\documentstyle[11pt]{article}
\input epsf.tex
\normalbaselines
\begin{document}
\bibliographystyle{unsrt}

\def\bea*{\begin{eqnarray*}}
\def\eea*{\end{eqnarray*}}
\def\ba{\begin{array}}
\def\ea{\end{array}}
\count1=1
\def\be{\ifnum \count1=0 $$ \else \begin{equation}\fi}
\def\ee{\ifnum\count1=0 $$ \else \end{equation}\fi}
\def\ele(#1){\ifnum\count1=0 \eqno({\bf #1}) $$ \else
\label{#1}\end{equation}\fi}
\def\req(#1){\ifnum\count1=0 {\bf #1}\else \ref{#1}\fi}
\def\bea(#1){\ifnum \count1=0   $$ \begin{array}{#1}
\else \begin{equation} \begin{array}{#1} \fi}
\def\eea{\ifnum \count1=0 \end{array} $$
\else  \end{array}\end{equation}\fi}
\def\elea(#1){\ifnum \count1=0 \end{array}\label{#1}\eqno({\bf #1}) $$
\else\end{array}\label{#1}\end{equation}\fi}
\def\cit(#1){
\ifnum\count1=0 {\bf #1} \cite{#1} \else
\cite{#1}\fi}
\def\bibit(#1){\ifnum\count1=0 \bibitem{#1} [#1    ] \else \bibitem{#1}\fi}
\def\ds{\displaystyle}
\def\hb{\hfill\break}
\def\comment#1{\hb {***** {\em #1} *****}\hb }

\newcommand{\ZZ}{\hbox{Z\hspace{-3pt}Z}}
\newcommand{\NZ}{\hbox{I\hspace{-2pt}N}}
\newcommand{\RZ}{\hbox{I\hspace{-2pt}R}}
\newcommand{\CZ}{\,\hbox{I\hspace{-6pt}C}}
\newcommand{\PZ}{\hbox{I\hspace{-2pt}P}}
\newcommand{\HZ}{\hbox{I\hspace{-2pt}H}}

\vbox{\vspace{38mm}}
\begin{center}
{\LARGE \bf Bethe Ansatz for Heisenberg XXX Model}\\[5mm]
Shao-shiung Lin \\
{\it Department of Mathematics,\\
Taiwan University \\ Taipei, Taiwan \\
 (e-mail: lin@math.ntu.edu.tw)} \\ [3 mm]
Shi-shyr Roan\footnote{Supported in part by
the NSC grant of Taiwan.}\\{\it Institute of Mathematics \\ Academia Sinica \\
Taipei , Taiwan \\ (e-mail: maroan@ccvax.sinica.edu.tw)} \\[5mm]
\end{center}

\begin{abstract}
We investigate Bethe Ansatz equations for the one-dimensional
spin-$\frac{1}{2}$ Heisenberg XXX chain with a special
interest in a finite system. Solutions for
the two-particle sector are obtained. The
ground state in antiferromagnetic case has been analytically
studied through the logarithmic form of Bethe Ansatz
equations.
\end{abstract}

\vfill
\eject

\section{Introduction}
This paper is devoted to the study of Bethe Ansatz equations (BAE) for
the isotropic spin-$\frac{1}{2}$ Heisenberg chain.
The model describes interacting spins, situated on the sites
of a periodic lattice $N$.
There has been a profound development in physical interest
on the thermodynamical properties of their solutions. In
recent years a strong increase of mathematical attention on
this subject arises from
the study of integrable systems via quantum inverse scattering
method.
In most cases the results obtained are mainly concerned with
the lattice size $N$ tending to infinity.
However the finite-size problems, or the finite-size
corrections,
should be of mathematical interest, as well as
essential to the understanding of BAE, which is the basis for
all exact solutions of electronic models in one dimension.
We intend to investigate here the finite-size BAE for Heisenberg
XXX model in a rigorous
mathematical way, which will serve as a standard theory
for our subsequent works on quantum integrable systems.

In his remarkable works \cite{B}, Baxter discovered a link
between the quantum spin models and transfer matrices of
2-dimensional statistical lattice models. The quantum inverse
scattering method \cite{TF79} provides a natural understanding
of the role of Bathe Ansatz in the problem of spectrum of
model Hamiltonians. For Heisenberg XXX Hamiltonian,
Bethe vectors have some special characteristic properties:
they are eigenvectors of of the corresponding transfer
matrices with polynomials as the eigenvalues, and also
the highest weight vectors for the global $sl_2$-symmetry of
the Hamiltonian. In the
investigation of solutions of BAE in the large $N$ limit,
there is one fundamental assumption, the
so-called string hypothesis (Takahashi \cite{Ta}).
We observe that a certain feature of these string structures
can be understood through BAE, while
the counting of states \cite{TF81} based on this string
hypothesis remains still valid for the 2-particle sector of
any finite system $N$. The analytical solutions for sectors other
than two particles are hard to obtain except a very small size,
e.g. $N=6$. However for the ground state, one can
study the problem by analysing the corresponding
logarithmic form equation,
where the fixed point theory can be used for the existence
of solutions. The uniqueness of the ground state solution
should be mathematically expected, but only strong indication
can be obtained here.
Such program is now under progress and partial results are
promising.

The organization of this paper is as follows. In Section 2,
we recall necessary definitions in the theory of quantum integrable
systems \cite{FT} and
give various
characterizations of Bethe roots for Heisenberg XXX model.
In Section 3, Bethe roots as $N \rightarrow \infty$ are
discussed, and the string structure for a large $N$ is derived
from BAE.
In Section 4, we study some problems on
 Heisenberg XXX model of a finite lattice size.
Here the BAE for the two-particle sector is solved
analytically, and
mathematical structures of BAE
on the ground
state in antiferromagnetic case, as well as the
equivalent equation for its logarithmic form, are discussed.
We establish rigorously the existence of ground
state via the fixed point theory, and also
the uniqueness of the solution for a small $N$.
In Section 5, we present an illuminating ( but not a
mathematically rigorous )
argument on the uniqueness of ground state for a large
finite system.
We have written this note
in a mathematical style, and hope that in process it
would not be much difficult to read for both mathematicians
and theoretical physicians.

\section{Characterizations of Bethe States}
The Hamiltonian of Heisenberg XXX model is given by
\[
H_{\rm XXX} = - \frac{J}{4} \sum_{n=1}^N ( \sigma_n^1\sigma_{n+1}^1 +
\sigma_n^2\sigma_{n+1}^2 + \sigma_n^3\sigma_{n+1}^3 - 1 ) \ ,
\ \ \ \ J \in \RZ \ .
\]
Here $\sigma^j$'s are Pauli matrices.
In this paper, we shall always assume the size $N$ to be
even with the periodic boundary condition ( $N+1 \equiv 1 $) .
The operator $H_{\rm XXX}$ acts on the Hilbert space of
physical states $\HZ_N$,
\[
\HZ_N = \bigotimes_{n=1}^N h_n \ , \ \ \ \ h_n : = \CZ^2 \ \mbox{
for \ all \ } \ \ n .
\]
The link between the above quantum XXX system
and a 2-dimensional statistical lattice model is described
as follows \cite{B} \cite{FT}. Define the
operators of $\CZ^2 \otimes h$
$$
\begin{array}{ll}
R ( \lambda )  = &  \left( \begin{array}{cc}
          \lambda \sigma^4 + \frac{i}{2} \sigma^3 &
         i \sigma^-  \\
          i \sigma^+  &
          \lambda \sigma^4 - \frac{i}{2} \sigma^3
         \end{array}   \right) \ , \ \ \ \  \sigma^{\pm} =
\frac{\sigma^1 \pm i \sigma^2}{2} \ , \ \lambda \in \CZ \ ,
\end{array}
$$
with the first factor $\CZ^2$
as the auxiliary space, and the second factor $h$ as the quantum
space with the basis
\[
|+> = \left( \begin{array}{c}
1 \\ 0
\end{array} \right) \ , \ \
|-> = \left( \begin{array}{c}
0 \\ 1
\end{array} \right) \ .
\]
The $R ( \lambda )$ satisfy the following
Yang-Baxter equation:
\[
{\cal R} ( \lambda - \mu ) ( R( \lambda ) \otimes R ( \mu )) =
( R( \mu ) \otimes R( \lambda ) ) {\cal R}( \lambda - \mu ) ,
\]
where ${\cal R} ( \lambda )$ is the $4 \times 4$ numerical
matrix defined by
$$
\begin{array}{ll}
{\cal R} ( \lambda ) = & \left( \begin{array}{cccc}
    1 & 0 & 0 & 0 \\
     0& {\bf b} ( \lambda )  & {\bf c} ( \lambda ) & 0 \\
0 & {\bf c} ( \lambda ) & {\bf b} ( \lambda ) &0 \\
0 & 0 & 0 & 1
\end{array} \right) , \ \ {\bf b} ( \lambda ) = \frac{i}{
\lambda + i } \ , \ \ {\bf c} ( \lambda ) = \frac{ \lambda}{
\lambda + i} \ .
\end{array}
$$
Using $R( \lambda )$, we introduce the local transform matrix
\[
L_n( \lambda ) = 1 \otimes \cdots \otimes R (\lambda)_{n th}
\otimes \cdots \otimes 1
\]
as the operators of $\CZ^2 \otimes \HZ_N$
having $R( \lambda )$ at the $n$-th side.
Define the monodromy matrix
\[
F_N ( \lambda ) = L_N ( \lambda ) L_{N-1} ( \lambda ) \cdots
L_1 ( \lambda ) =
\left( \begin{array}{cc}
    A ( \lambda ) & B ( \lambda ) \\
    C ( \lambda ) & D ( \lambda )
\end{array} \right) .
\]
Then we have
\[
{\cal R}( \lambda - \mu ) ( F_N( \lambda ) \otimes F_N ( \mu ) ) =
( F_N ( \mu ) \otimes F_N( \lambda ) ) {\cal R}( \lambda - \mu ) .
\]
The matrix entry elements
$A ( \lambda ), B ( \lambda ), C ( \lambda ) , D ( \lambda )$
of the monodromy $F ( \lambda )$
are operators of $\HZ_N$, which
generate the so called ABCD algebra:
\bea(cll)
A ( \lambda ) A ( \mu ) = A ( \mu ) A ( \lambda ) , &
B ( \lambda ) B ( \mu ) = B ( \mu ) B ( \lambda ) , & \\
C ( \lambda ) C ( \mu ) = C ( \mu ) C ( \lambda ) , &
D ( \lambda ) D ( \mu ) = D ( \mu ) D ( \lambda ) , & \\
A ( \mu ) B ( \lambda ) = f_{ \mu , \lambda } B ( \lambda ) A ( \mu )
& + \ g_{ \lambda , \mu } B ( \mu ) A ( \lambda ) ,
& A \leftrightarrow B , \\
D ( \lambda ) B ( \mu ) = f_{ \mu , \lambda  } B ( \mu )
D ( \lambda )
 & + \ g_{  \lambda, \mu } B ( \lambda ) D ( \mu ) ,&
B \leftrightarrow D , \\
C ( \lambda ) A ( \mu ) = f_{ \mu , \lambda } A ( \mu ) C ( \lambda )
 & + \ g_{ \lambda , \mu } A ( \lambda ) C ( \mu ) ,&
A \leftrightarrow C   \\
C ( \mu ) D ( \lambda ) = f_{ \mu , \lambda  } D ( \lambda )
C ( \mu ) & + \ g_{ \lambda , \mu } D ( \mu ) C ( \lambda ) ,
& C \leftrightarrow D   \\
C ( \lambda )B ( \mu ) - B ( \mu ) C ( \lambda )  = &g_{ \lambda , \mu }
( A ( \lambda ) D ( \mu ) - A ( \mu ) D ( \lambda ) )  =&
 g_{ \mu , \lambda  } (
D( \lambda ) A ( \mu ) - D ( \mu ) A ( \lambda ) ) , \\
D ( \lambda )A ( \mu ) - A ( \mu ) D ( \lambda )  =& g_{ \lambda , \mu }
( B ( \lambda ) C ( \mu ) - B ( \mu ) C ( \lambda ) )  =&
 g_{ \mu , \lambda  } (
C( \lambda ) B ( \mu ) - C ( \mu ) B ( \lambda )   )  .
\elea(ABCD)
where
\[
f_{ \mu, \lambda  } = \frac{1}{ {\bf c} ( \lambda - \mu ) }
 , \ \ \
g_{ \lambda , \mu } = - \frac{ {\bf b} ( \lambda - \mu ) }
{ {\bf c} (
\lambda - \mu ) }  .
\]
Taking the trace of the monodromy, one obtains the
transfer matrices
\[
T_N ( \lambda ) := \mbox{tr} F_N ( \lambda ) = A (\lambda ) + D ( \lambda )
\ ,
\]
which form a family of commuting operators of $\HZ_N$:
\[
T_N ( \lambda ) \cdot T_N ( \mu ) = T_N ( \mu ) \cdot T_N ( \lambda ) .
\]
The Hamiltonian
of XXX model is related to the transfer matrices by
\[
H_{\rm XXX} = -\frac{J}{2} (i \frac{d}{d\lambda}
\log T( \lambda )|_{ \lambda = \frac{i}{2} } - N ) \ \ .
\]
Define the pseudo-vacuum
\[
\Omega_N := \bigotimes_{n=1}^N v_n \in \HZ_N \ , \ \ \ \ v_n = |+>
\ \ \mbox{for \ all \ } \ n .
\]
Then we have
\bea(c)
A ( \lambda ) \Omega_N = (\lambda + \frac{i}{2})^N  \Omega_N
\ , \ \ \
D ( \lambda ) \Omega_N = (\lambda - \frac{i}{2})^N  \Omega_N
\ , \ \ \
C ( \lambda ) \Omega_N = 0 \ , \\ [4mm]
B ( \lambda ) \Omega_N = i \sum_{n=1}^N (\lambda + \frac{i}{2} )^{n-1}
(\lambda - \frac{i}{2} )^{N - n}
(1 \otimes \cdots \otimes \sigma^{-}_{n th}
\otimes \cdots \otimes 1) \Omega_N \ .
\elea(ADCBvacuum)
For complex numbers $\lambda_j, 1 \leq j \leq l ,$ we consider
the vector
\[
\Psi_N ( \lambda_1, \ldots , \lambda_l ) : =
\prod_{ m = 1}^l B ( \lambda_m ) \Omega_N \ \in \ \HZ_N \ ,
\]
and define the function of $\lambda$,
\be
\Lambda ( \lambda ; \lambda_1, \ldots , \lambda_l ) : =
(\lambda + \frac{i}{2})^N  \prod_{ m =1}^l
\frac{ \lambda - \lambda_m - i }{ \lambda - \lambda_m }
 +
(\lambda - \frac{i}{2})^N  \prod_{ m =1}^l
\frac{ \lambda - \lambda_m + i }{ \lambda - \lambda_m } \ .
\ele(Lambda)
One has
\[
\Lambda ( \overline{\lambda} ; \overline{ \lambda_1}, \ldots ,
\overline{ \lambda_l} ) =
\overline{\Lambda ( \lambda ; \lambda_1, \ldots ,
\lambda_l )} \ , \ \ \
\Lambda ( - \lambda ; - \lambda_1 , \ldots , - \lambda_l ) =
\Lambda ( \lambda ; \lambda_1 , \ldots , \lambda_l) \ ,
\]
hence
$$
\begin{array}{lll}
\{ \lambda_j \}_j \subseteq \RZ \ & \Longleftrightarrow &
\overline{\Lambda ( \lambda ; \lambda_j)} =
\Lambda ( \overline{\lambda} ; \lambda_j) \ \\
\{ \lambda_j \}_j = \{ - \lambda_j \}_j & \Longleftrightarrow &
\Lambda ( - \lambda ; \lambda_j) =
\Lambda ( \lambda ; \lambda_j) \ .
\end{array}
$$
By the relations
\[
\prod_{ m =1}^l
\frac{ \lambda - \lambda_m - i }{ \lambda - \lambda_m }
= 1 + \sum_{m=1}^l
\frac{ a_m }{ \lambda - \lambda_m }  \ , \ \ \ \
\prod_{ m =1}^l
\frac{ \lambda - \lambda_m + i }{ \lambda - \lambda_m }
= 1 + \sum_{m=1}^l
\frac{ b_m }{ \lambda - \lambda_m }
\]
where
\[
a_m = -i \prod_{ j \neq m} ( 1 - \frac{ i }{ \lambda_m -
\lambda_j } ) \ , \ \
b_m = i \prod_{ j \neq m} ( 1 + \frac{ i }{ \lambda_m -
\lambda_j } ) \ ,
\]
one has
\be
\Lambda ( \lambda ;
\lambda_1, \ldots , \lambda_l ) =
( \lambda + \frac{i}{2} )^N
+ ( \lambda - \frac{i}{2} )^N  +
\sum_{m=1}^l
\frac{ a_m ( \lambda + \frac{i}{2} )^N  +
b_m ( \lambda - \frac{i}{2} )^N }{ \lambda - \lambda_m } \ .
\ele(ab)
The criterion for $\Lambda ( \lambda ; \lambda_1, \ldots ,
\lambda_l ) $ to be an entire function of $\lambda$ is now
equivalent to $\lambda_j$'s satisfying the
Bethe Ansatz equation ( BAE ) :
\be
( \frac{ \lambda_j + \frac{i}{2} }{ \lambda_j - \frac{i}{2} } )^N =
\prod_{m=1, m \neq j}^l \frac{ \lambda_j - \lambda_m + i
}{ \lambda_j - \lambda_m - i} \ \ , \ \ \ \lambda_j \in \CZ , \ \
j = 1 , \ldots, l .
\ele(BAE)
In this situation, the difference of $\Lambda ( \lambda ;
\lambda_1, \ldots , \lambda_l ) $
and $( \lambda + \frac{i}{2} )^N +
( \lambda - \frac{i}{2} )^N $ is a polynomial of degree at most
$N-1$.
Note that $( \lambda + \frac{i}{2} )^N
+ ( \lambda - \frac{i}{2} )^N $ is the eigenvalue of
the transfer matrices $T_N ( \lambda )$ for
the pseudo-vacuum ,
\[
 T_N ( \lambda )  \Omega_N  = (( \lambda + \frac{i}{2} )^N
+ ( \lambda - \frac{i}{2} )^N ) \Omega_N \ .
\]
The variable $\lambda$ can be interpreted as the
"rapidity" of a "particle" with its momentum $p ( \lambda )$
defined by
\[
 e^{ip( \lambda )} = \frac{ \lambda + \frac{i}{2}}
{\lambda - \frac{i}{2} } \  \]
or equivalently ,
\[
\lambda = \cot \frac{ p( \lambda ) }{2} \ .
\]
The scattering amplitude
$S ( \lambda , \mu )$ of two particles is
\[
e^{i\phi( \mu , \lambda )}  = S ( \lambda , \mu )
 =  \frac{ \mu - \lambda  + i }{ \mu -
\lambda  - i }
\]
with the scattering angle satisfying the relation
\[
2 \cot \frac{ \phi ( \mu , \lambda ) }{2} =
\cot \frac{ p( \mu ) }{2} - \cot \frac{ p( \lambda ) }{2} \ .
\]
Now BAE (\req(BAE)) also takes the form
\[
 e^{ip( \lambda_k )N} =  - \prod_{ m=1 }^l
e^{i\phi( \lambda_k , \lambda_m )} \ \ , \ \ \ \ k = 1, \ldots , l .
\]
The right hand side is the $l$-particle scattering amplitude
in terms of two-particle ones, a fact that manifests the
integrability of the model. By the relation
\[
S ( \lambda , \mu ) S ( \mu , \lambda ) = 1 \ ,
\]
one obtains
\be
 \prod_{ k = 1 }^l e^{ip( \lambda_k )N} =
\prod_{ k = 1 }^l
( \frac{ \lambda_k + \frac{i}{2}}{\lambda_k - \frac{i}{2} })^N
= 1 \ .
\ele(constotalmo)
The vector $\Psi_N ( \lambda_1, \ldots , \lambda_l ) $ is
called a Bethe vector when $\lambda_j$'s form a solution of BAE.

Denote
\[
S_j = \frac{1}{2} \sum_{n=1}^N \sigma_n^j \ \ \  \ ( j = 1 , 2
, 3 ) \ ; \ \ \ \ S_{\pm} = S_1 \pm i S_2 \ .
\]
$S_{\pm} , S_3$ form a basis of $sl_2( \CZ )$ acting
on $\HZ_N$. One has
$$
\begin{array}{rll}
[ L_n ( \lambda ) , S_j ] =& \frac{i}{4}  \sum_{k=1}^3
[ \sigma^k \otimes \sigma_n^k , \sigma_n^j ]
=& \frac{-1}{2}  \sum_{k , l =1}^3 \epsilon_{kjl}
 \sigma^k \otimes \sigma_n^l  \\[2mm]
=& \frac{1}{2}  \sum_{k , l =1}^3 \epsilon_{ljk}
 \sigma^k \otimes \sigma_n^l
=& \frac{-i}{4}  \sum_{ l =1}^3
 [ \sigma^l ,  \sigma^j ] \otimes \sigma_n^l  \\[2mm]
=& \frac{i}{4}  \sum_{ l =1}^3
 [ \sigma^j , \sigma^l \otimes \sigma_n^l ]_{aux}
=& \frac{1}{2}  [ \sigma^j , L_n ( \lambda ) ]_{aux}
\end{array}
$$
here the Lie-product $[M , b ]$ of a $2 \times 2$ matrix $M$,
\[
M =  \left( \begin{array}{cc}
  M_{11} &  M_{12} \\
 M_{21} &  M_{22}
\end{array} \right)
\]
with an element $b$ in an arbitrary ring is defined to be the
$2 \times 2$ matrix
\[
[ M , b ] = \left( \begin{array}{cc}
 [ M_{11}, b ] & [ M_{12}, b ] \\

[ M_{21}, b ] & [ M_{22}, b]
\end{array} \right) \ ,
\]
and the Lie-bracket of the last term is on the auxiliary
space. It follows
\[
[ F_N ( \lambda ) , S_j ] =  \frac{1}{2}
[ \sigma^j , F_N ( \lambda ) ]_{aux}
\]
and
$$
\begin{array}{c}
[ A ( \lambda ) , S_3 ] = [ D ( \lambda ) , S_3 ]
 = [ C ( \lambda ) , S_+ ] = [ C ( \lambda ) , S_- ]
 = 0 , \\[2mm]
[ B ( \lambda ) , S_3 ] = B ( \lambda ) \ , \ \ \
[ C ( \lambda ) , S_3 ] = - C ( \lambda ) , \\[2mm]
[ B ( \lambda ) , S_+ ] =
- [ C ( \lambda ) , S_- ] = - A ( \lambda ) + D ( \lambda ) ,
\\[2mm]
[ A ( \lambda ) , S_+ ] = - [ D ( \lambda ) , S_+ ] = C ( \lambda )
\\[2mm]
[ A ( \lambda ) , S_- ] = - [ D ( \lambda ) , S_- ] = - B ( \lambda )
\end{array}
$$
Therefore
\[
[ T_N ( \lambda ), S_j ] = 0 \ ,
\]
which states the $sl_2$-symmetry property of the Hamiltonian
$H_{\rm XXX}$. It is easy to see that
\[
S_{+} \Omega_N = 0 \ \ \ , \ \ \ \ \
S_3 \Omega_N = \frac{N}{2} \Omega_N
\]
and the spin of $\Psi_N ( \lambda_1, \cdots , \lambda_l )$
is $\frac{N}{2} -l$ ,
\[
S_3 \Psi_N ( \lambda_1, \cdots , \lambda_l ) = ( \frac{N}{2}
-l) \Psi_N ( \lambda_1, \cdots , \lambda_l ) .
\]
The following  conditions for a
Bethe vector should be well-known specialists, but we could not
find explicit references. Here we give the details of the proof.
\par \vspace{0.2mm} \noindent
{\bf Proposition 1 .}
$\Psi_N ( \lambda_1, \ldots , \lambda_l ) $ is a Bethe vector
if and only if one of
the following equivalent conditions holds :

(i) $\{ \lambda_j \}_{j=1}^l$ satisfies BAE.

(ii) The function $\Lambda ( \lambda ; \lambda_1, \ldots ,
\lambda_l ) $ has no pole at a finite value of $\lambda$.

(iii) $\Psi_N ( \lambda_1, \ldots , \lambda_l ) $
is a common eigenvector of transfer matrices $T_N ( \lambda )$.

(iv) $\Psi_N( \lambda_1, \ldots , \lambda_l )$ is a highest
weight vector ( of spin $\frac{N}{2} -l$ )
for the $sl_2$-representation.
\par \vspace{0.2mm} \noindent
{\it Proof. } The equivalence of (i) and (ii) is known before.

(i) $\leftrightarrow$ (iii).
By the relations (\req(ABCD))
(\req(ADCBvacuum)), one can
move $A ( \lambda )$ and $ D ( \lambda )$ through $B ( \lambda_m )$
to $\Omega_N$ in the expression
\[
T_N( \lambda ) \prod_{m=1}^l B( \lambda_m ) \Omega_N \
= ( A ( \lambda ) + D ( \lambda ) ) \prod_{m=1}^l B( \lambda_m )
\Omega_N \ .
\]
The resulting form becomes a combination of
\[
\Lambda ( \lambda ; \lambda_1, \ldots , \lambda_l )
\prod_{m=1}^l B( \lambda_m ) \Omega_N
\]
with the forms
\[
\Lambda_k ( \lambda ; \lambda_1, \ldots , \lambda_l )
B( \lambda ) \prod_{m=1, m \neq k}^l
B( \lambda_m ) \Omega_N \ , \ \ ( k = 1, \ldots , l ) .
\]
The expression of $\Lambda ( \lambda ;
\lambda_1, \ldots , \lambda_l )$ is given by (\req(Lambda))
by taking
account only the first terms on the right hand side of
the commutation relations of $A , B$ and $D, B$ in (\req(ABCD)).
If we use second terms of (\req(ABCD)) on the commutation
relations of $A( \lambda )$ , $D ( \lambda )$ with
$B ( \lambda_1 )$ , and then first terms of (\req(ABCD))
on the commutation relations
of $A( \lambda_1 )$ , $D ( \lambda_1 )$ with
$B ( \lambda_m ) $ for $ m \geq 2$, by the relation
$g_{ \lambda , \mu } = -  g_{ \mu , \lambda }$ ,
we obtain the expression of $\Lambda_1 ( \lambda ;
\lambda_1, \ldots , \lambda_l )$:
\[
\Lambda_1 ( \lambda ; \lambda_1, \ldots , \lambda_l )
= g_{ \lambda_1, \lambda } [ ( \lambda_1 + \frac{i}{2} )^N
\prod_{m=2}^l f_{ \lambda_1, \lambda_m}
  - ( \lambda_1 - \frac{i}{2}  )^N
\prod_{m=2}^l f_{ \lambda_m, \lambda_1} ] \ .
\]
Since $B ( \lambda_m )$'s are commuting operators, by a suitable
permutation of the indices $m$, one concludes the
expression of
$\Lambda_k ( \lambda ; \lambda_1, \ldots , \lambda_l )$
which is given by
\[
\Lambda_k ( \lambda ; \lambda_1, \ldots , \lambda_l )
= g_{ \lambda_k, \lambda } [ ( \lambda_k + \frac{i}{2}  )^N
\prod_{m=1, m \neq k}^l
f_{ \lambda_k, \lambda_m}
 -  ( \lambda_k - \frac{i}{2} )^N
\prod_{m=1, m \neq k}^l f_{ \lambda_m, \lambda_k} ] \ .
\]
Therefore $
\Psi_N ( \lambda_1, \ldots , \lambda_l )$ is an eigenvector
of $T_N ( \lambda )$ provided $\{ \lambda_k \}_{k=1}^l$
satisfies the relations
\[
\Lambda_k ( \lambda ; \lambda_1, \ldots , \lambda_l ) = 0 \ ,
\ \  \mbox{ \ for \ }
k = 1 , \ldots , l , \mbox{ \ and \ all \ } \lambda ,
\]
which is equivalent to BAE.
Therefore we obtain the result.

(i) $\leftrightarrow$ (iv).
Using the relations between $B( \lambda )$ and
$S_j$, one has
\[
S_+ \Psi_N ( \lambda_1, \cdots , \lambda_l ) =
\sum_{j=1}^lB( \lambda_1 ) \cdots B( \lambda_{j-1}) ( A
( \lambda_j) - D ( \lambda_j )) B( \lambda_{j+1})
\cdots B( \lambda_l ) \Omega_N .
\]
By using (\req(ABCD)),
$S_+ \Psi_N ( \lambda_1, \cdots , \lambda_l )$ can be
expressed as a combination of forms
\[
M_k ( \lambda_1, \ldots, \lambda_l ) B ( \lambda_1) \cdots
B( \lambda_{k-1}) B ( \lambda_{k+1}) \cdots B( \lambda_l )
\Omega_N \ , \ \ \ ( k = 1 , \ldots , l ) .
\]
Taking account only of first terms on the right hand side of
the commutation relations between $A , B$ and $D, B$ in
(\req(ABCD)), one obtains
\[
M_1 ( \lambda_1, \ldots, \lambda_l ) = ( \lambda_1 +
\frac{i}{2} )^N f_{ \lambda_1, \lambda_2 }
f_{ \lambda_1, \lambda_3 } \ldots f_{ \lambda_1,
\lambda_l }  - ( \lambda_1 -
\frac{i}{2} )^N f_{ \lambda_2, \lambda_1 }
f_{ \lambda_3, \lambda_1 } \ldots
f_{ \lambda_l, \lambda_1 } .
\]
Since $B( \lambda_m)$'s are commuting operators, by the symmetry
argument, one also has
\[
M_k ( \lambda_1, \ldots, \lambda_l ) = ( \lambda_k +
\frac{i}{2}  )^N \prod_{m \neq k}^l
f_{ \lambda_k, \lambda_m }  - ( \lambda_k - \frac{i}{2}  )^N
\prod_{m \neq k}^lf_{ \lambda_m, \lambda_k }.
\]
The vanishing of $M_k ( \lambda_1, \ldots, \lambda_l )$ is now
equivalent to BAE. Hence we obtain the equivalence
between (i) and (iv).
$\Box$ \par \vspace{0.2mm} \noindent
{\bf Remark.} The Bethe states are in an one-one correspondence
with the solutions of BAE. Indeed one has the following
equivalent statements for Bethe vectors:
$$
\begin{array}{ll}
& \Psi_N( \lambda_1, \cdots , \lambda_l ) = s
\Psi_N( \lambda_1^\prime, \cdots , \lambda_k^\prime )
\mbox{\ for\ some\ } s \in \CZ - \{ 0 \}   \\
\Longleftrightarrow &
\Lambda ( \lambda ;
\lambda_1, \ldots , \lambda_l ) = \Lambda ( \lambda ;
\lambda_1^\prime, \ldots , \lambda_k^\prime )  \\
\Longleftrightarrow &
l = k \ \ \mbox{and} \ \ \ \{ \lambda_1, \ldots , \lambda_l \} =
\{ \lambda_1^\prime, \ldots , \lambda_k^\prime \} \ .
\end{array}
$$
It is obvious that two linear dependent Bethe vectors
$\Psi_N( \lambda_1, \cdots , \lambda_l )$ and
$\Psi_N( \lambda_1^\prime, \cdots , \lambda_k^\prime )$
 have the same eigenvalues:
\[
\Lambda ( \lambda ;
\lambda_1, \ldots , \lambda_l ) = \Lambda ( \lambda ;
\lambda_1^\prime, \ldots , \lambda_k^\prime ) ,
\]
and by (\req(ab)), this means
\[
\sum_{m=1}^l
\frac{ a_m ( \lambda + \frac{i}{2} )^N +
b_m ( \lambda - \frac{i}{2} )^N }{ \lambda - \lambda_m}
= \sum_{j =1}^k
\frac{ a_j^\prime ( \lambda + \frac{i}{2} )^N +
b_j^\prime ( \lambda - \frac{i}{2} )^N }{ \lambda -
\lambda_j^\prime } \ ,
\]
where
$$
 \begin{array}{ll}
a_m = -i \prod_{ j \neq m} ( 1 - \frac{ i }{ \lambda_m -
\lambda_j } ) \ \ , &
b_m = i \prod_{ j \neq m} ( 1 + \frac{ i }{ \lambda_m -
\lambda_j } ) \ , \\
a_m^\prime = -i \prod_{ j \neq m} ( 1 - \frac{ i }
{ \lambda^\prime_m -
\lambda_j^\prime } ) \ \ , &
b_m^\prime = i \prod_{ j \neq m} ( 1 + \frac{ i }{
\lambda^\prime_m - \lambda_j^\prime } ) \ .
\end{array}
$$
Note that by BAE,
$$
\begin{array}{lll}
a_m = 0  & \Longleftrightarrow \lambda_m = \frac{i}{2}  &
\Longrightarrow b_m \neq 0 \ ,
\\
b_m = 0 & \Longleftrightarrow \lambda_m = \frac{-i}{2} \ &
\Longrightarrow a_m \neq 0 \ .
\end{array}
$$
Then we have
\[
( \lambda + \frac{i}{2} )^N (
\sum_{m=1}^l \frac{ a_m }{ \lambda - \lambda_m} -
\sum_{j =1}^k
\frac{ a_j^\prime }{ \lambda - \lambda_j^\prime }) =
- ( \lambda - \frac{i}{2} )^N  (
\sum_{m=1}^l \frac{ b_m }{ \lambda - \lambda_m} -
\sum_{j =1}^k
\frac{ b_j^\prime }{ \lambda - \lambda_j^\prime }) \ .
\]
By multipling $\prod_{m=1}^l(\lambda - \lambda_m)
\prod_{j=1}^k(\lambda - \lambda_j^\prime)$ on the
above equation and from $l + k \leq N$, we have
\[
\sum_{m=1}^l \frac{ a_m }{ \lambda - \lambda_m} -
\sum_{j =1}^k
\frac{ a_j^\prime }{ \lambda - \lambda_j^\prime } \equiv 0 \ ,
\ \ \ \
\sum_{m=1}^l \frac{ b_m }{ \lambda - \lambda_m} -
\sum_{j =1}^k
\frac{ b_j^\prime }{ \lambda - \lambda_j^\prime } \equiv 0 \ .
\]
This implies that each
$\lambda_m$ is equal to some $\lambda_j^\prime$. Similarly
$\lambda_j^\prime$ is equal to $\lambda_m$ for some $m$,
hence
\[
l = k \ \ \mbox{and} \ \ \ \{ \lambda_1, \ldots , \lambda_l \} =
\{ \lambda_1^\prime, \ldots , \lambda_k^\prime \} \ .
\]
So we obtain the conclusions.
$\Box$ \par \vspace{0.2mm} \noindent

\section{String Hypothesis of Roots of Bethe Ansatz Equation}
In the description of solutions of BAE as the size
$N$ tends to infinity, one assumes the string hypothesis
\cite{Ta} which claims any solution consists of a series
of strings in the form
\[
\lambda = x + i ( \frac{n+1}{2} - k ) + O ( e^{ - \beta N } )
\ \ \ \  k = 1 \ldots n  \ \ \ \ \ x \in \RZ ,
\]
for some $\beta > 0$. In this section we shall discuss the
relation between the string structure and BAE.
First we define certain notions needed for our
purpose.
A complex number $z$ is called an asymptotic limit point of
a sequence of Bethe roots
$\{ \lambda_j^{(N)} \}_{j=1}^{l(N)}$ if for some choice of
$j (N)$, the following relation holds:
\[
 \lambda_{j(N)}^{(N)} = z +
O ( e^{ - \beta N } ) \ \ \ \
{\rm as \ } \ N \rightarrow \infty \ .
\]
For the convenience, we shall simply write
\[
\lambda_j^{N} \sim z  \ \ \ \
{\rm as} \ \ N \rightarrow \infty
\]
when no confusion could arise. When the asymptotic limit point
$z$ is non-real, i.e. ${\rm Im} (z) \neq 0$, the above element
$ \lambda_{j(N)}^{(N)}$ is called a complex
root in the Bethe solution $\{ \lambda_j^{(N)} \}_{j=1}^{l(N)}$.
Moreover, if there exists another sequence
$ \lambda_{k(N)}^{(N)}$ whose asymptotic limit
point is the real part of a
complex root, $\lambda_{k(N)}^{(N)}$ is also defined to be a
complex root.

A more precise description for string hypothesis states as
follows: A Bethe solution for
a large $N$ always lies in a collection of Bethe
solutions $\{ \lambda_j^{(N)} \}_{j=1}^{l(N)}, ( \ N \gg 0 ),$ such that
every root belongs to a suitable convergent sequence
of the form:
\[
 \lambda_j^{(N)} \sim x + i ( \frac{n+1}{2} - k )  \ \ \ \
{\rm as \ } \ N \rightarrow \infty \ .
\]
The collection
\be
x + i ( \frac{n-1}{2} - j ) \ \ , \ \ 0 \leq j \leq n-1 \ \ \
\ \ x \in \RZ \ ,
\ele(string)
will be called a string of length $n$ with center $x$.
We shall derive this string structure
of Bethe roots as $N$ tends to infinty under the following
additional ( somewhat unpleasant ) condition:

Hypothsis (H) : " There exists a positive integer $N_0$ such that any
Bethe root $\{ \lambda_j \}_{j=1}^l$ for
a large size always lies in a sequence of Bethe
solutions $\{ \lambda_j^{(N)} \}_{j=1}^{l(N)},
N \geq N_0$,
which tends to an asymptotic configuation consisting of elements
in real axis with a finite number of complex numbers.
Futhermore the number of
complex roots of Bethe solutions remains constant in the limit
process as $N \rightarrow \infty$."

 First, let us consider
the simplest case for $l = 1$. BAE is given by
\be
( \frac{ \lambda + i/2}{ \lambda - i/2 } )^N = 1 \ , \ \
\lambda \in \CZ \ .
\ele(BAE1)
As $N$ tends infinity, the above equation becomes
\[
| \frac{ \lambda + i/2}{ \lambda - i/2 } | = 1 \ ,
\]
which simply means the momentum $p ( \lambda )$ being real,
i.e. the real rapidity $\lambda$, and the Bethe
state $\Psi_N ( \lambda )$ is called a magnon state with
the energy given by $ \frac{J}{2 ( \lambda^2 + \frac{1}{4} )}$ .

For $l > 0$, we describe the following lemma which somewhat
suggests the conjujate symmetric nature of a string.
\par \vspace{0.2mm} \noindent
{\bf Lemma 1. } Let $\{ \lambda_j \}_{j=1}^l$ be a solution of BAE
with $N \rightarrow \infty$ . Assume
\[
\lambda_j = \left\{ \begin{array}{ll}
x + i ( y - j ) & j \leq n \\
\in \RZ & j > n
\end{array}  \right.
\]
for some positive integer $n$ and real numbers $x, y$ . Then
$\lambda_1, \ldots , \lambda_n$ form a string of length $n$ with center at
$x$.
\par \vspace{0.2mm} \noindent
{\it Proof.} We have
\[
\prod_{j=1}^n \frac{\lambda_j + \frac{i}{2}}
{\lambda_j - \frac{i}{2}} = \frac{x + i (y - \frac{1}{2}) }
{x + i (y - n - \frac{1}{2})} \ \ {\rm and } \  \ \
| \frac{\lambda_k + \frac{i}{2}}
{\lambda_k - \frac{i}{2}} | = 1 \ \ \mbox{for\ } k > n .
\]
By (\req(constotalmo)) and letting $N \rightarrow \infty$, one has
\[
| \frac{x + i (y - \frac{1}{2})}
{x + i (y - n - \frac{1}{2})} | = 1
\]
which implies $ y = \frac{n + 1}{2}$. Therefore $\{ \lambda_j
\}_{ j= 1}^n$ is a string of length $n$.
$\Box$ \par \vspace{0.2mm} \noindent

For $l = 2$, BAE is given by
\[
( \frac{ \lambda_1 + i/2}{ \lambda_1 - i/2 } )^N =
\frac{ \lambda_1 - \lambda_2 + i }{
\lambda_1 - \lambda_2 - i } \ \ , \ \ \
( \frac{ \lambda_2 + i/2}{ \lambda_2 - i/2 } )^N =
\frac{ \lambda_2 - \lambda_1 + i }{
\lambda_2 - \lambda_1 - i } \ .
\]
We have
\[
( \frac{ \lambda_1 + i/2}{ \lambda_1 - i/2 })^N
( \frac{ \lambda_2 + i/2}{ \lambda_2 - i/2 } )^N = 1 \ .
\]
Either both $\lambda_j$'s are real,
or both not. If $\lambda_1$ and $\lambda_2$ are real,
$\Psi_N ( \lambda_1 , \lambda_2 )$ is called a 2 magnon state
and its energy equals to the sum of those of
1 magnon states $\Psi_N ( \lambda_1 )$ and
$\Psi_N ( \lambda_2 )$.
In
the case for both $\lambda_1 , \lambda_2$ not real, we may assume
$| \frac{ \lambda_1 + i/2}{ \lambda_1 - i/2 } | $ is greater
than one, hence
$\lim_{N \rightarrow \infty}
| \frac{ \lambda_1 + i/2}{ \lambda_1 - i/2 } |^N = \infty$.
The first relation of BAE implies
$ \lambda_1 = \lambda_2 + i$. By Lemma 1 ,
 $\lambda_1$ and $ \lambda_2$ form a string of length 2,
hence $\lambda_1= \bar{ \lambda_2} = x + \frac{i}{2}
$ for some $x \in \RZ$. In this case
$\Psi_N ( \lambda_1, \lambda_2 )$ is called a
bounded state, and its energy is equal to
$\frac{J}{x^2 + 1}$ .

For $l = 3$, let $\{ \lambda_1 , \lambda_2 , \lambda_3 \}$ be
a solution of BAE. By the realtion
\[
\prod_{j=1}^3 ( \frac{ \lambda_j + \frac{i}{2} }
{ \lambda_j - \frac{i}{2} } )^N = 1 \ ,
\]
either all the $\lambda_j$'s are real, or at least two of them
are not real numbers. The former case is the 3 magnon
state $\Psi_N ( \lambda_1, \lambda_2 , \lambda_3 )$. Otherwise
we may assume $\lambda_1$ and $\lambda_2 $ are not real with
\[
| \frac{ \lambda_1 + \frac{i}{2} }
{ \lambda_1 - \frac{i}{2} } | > 1 \ , \ \
| \frac{ \lambda_2 + \frac{i}{2} }
{ \lambda_2 - \frac{i}{2} } | < 1 \ .
\]
By BAE for $\lambda_1 , \lambda_2$ with $N \rightarrow \infty$,
we have
\[
\left\{ \begin{array}{ll}
( \lambda_1 - \lambda_2 - i ) ( \lambda_1 - \lambda_3 - i ) &
= 0 \ , \\
( \lambda_2 - \lambda_1 + i ) ( \lambda_2 - \lambda_3 + i ) &
= 0 \ .
\end{array} \right.
\]
Either $\lambda_1 = \lambda_2 + i$,
or $ \lambda_3 =
\lambda_1 - i =   \lambda_2 + i $.
For $\lambda_3 \in \RZ$ , by Lemma 1, either
$\{ \lambda_1 , \lambda_2 \}$ forms a string of length 2, or
$\{ \lambda_2 , \lambda_3, \lambda_1 \}$ forms a string of length
three. For
$\lambda_3 \not\in \RZ$,
$| \frac{ \lambda_3 + \frac{i}{2} }
{ \lambda_3 - \frac{i}{2} } | \neq 1$. By BAE for $\lambda_3$
with large $N$ limit, one has
\[
\{ \lambda_1 , \lambda_2 , \lambda_3 \} =
\{ z - i , z , z + i \} \ \ \mbox{for \ some \ } z \in \CZ ,
\]
hence it is a string of length three by Lemma 1, which
contradicts $\lambda_j \not\in \RZ$ for
all $j$. In this way we have determined
the structure of $\{ \lambda_j \}_{j=1}^3$ as $N$ tends
infinity, which is composed
of either 3 real roots,
a real with a 2-string , or a string of length 3 .

For $l \geq 4$, the procedure we employ above is not
sufficient to derive the string structure of Bethe solutions,
e.g. one needs to exclude a chain like
\[
\lambda_1 = \frac{-3i}{4} \ , \ \
\lambda_2 = \frac{-i}{4} \ , \ \
\lambda_3 = \frac{i}{4} \ , \ \
\lambda_4 = \frac{3i}{4} \ ,
\]
in the $l=4$ solutions. Now
let $\{ \lambda_j \}_{j=1}^l$ be a set of Bethe roots. Assume
\be
\mu_k : = \mu - ( k - 1 ) i \in \{ \lambda_j \}_{j=1}^l \ , \
\ k = 1, \ldots, m \
\ele(muk)
where $\mu$ is a complex number and $m$ is a positive integer
greater than 1. Denote
\[
\{ \nu_h \}_{h=1}^{l-m} = \{ \lambda_j \}_{j=1}^l -
\{ \mu_k \}_{k=1}^m \ .
\]
By multiplying the equations for $\mu_k$ in BAE, we have
\be
  ( \frac{ \mu_1 + \frac{i}{2} }
{  \mu_m - \frac{i}{2}  } )^N =
\prod_{k=1}^m \prod_{h=1}^{l-m}
 \frac{ \mu_k - \nu_h + i
}{ \mu_k - \nu_h - i} \ .
\ele(BAEm)
{}From
\[
 \frac{ \mu_1 + \frac{i}{2} }
{  \mu_m - \frac{i}{2}  } = \frac{ \mu + \frac{i}{2} }
{  \mu - \frac{(2m-1)i}{2}  } \ ,
\]
one has
\[
| \frac{ \mu_1 + \frac{i}{2} }
{  \mu_m - \frac{i}{2}  }  | = 1 \ , \ \
> 1 \ , \ \ < 1
 \ \Longleftrightarrow  \ \ {\rm Im} ( \mu )
= \frac{m-1}{2} \ , \ \ > \frac{m-1}{2} \ , \ \
< \frac{m-1}{2}
\]
respectively. Since the right hand side of (\req(BAEm)) is equal
to
\[
\prod_{h=1}^{l-m}
 \frac{ ( \mu_1 - \nu_h + i ) (\mu_1 - \nu_h)
}{ ( \mu_m - \nu_h) ( \mu_m - \nu_h - i) }  \ ,
\]
from our assumption (H) one has
$$
\begin{array}{lll}
{\rm Im} ( \mu ) > \frac{m-1}{2} & \Longrightarrow&
\lim_{N \rightarrow \infty} \nu_h =  \mu_m - i
 \ \ \ \ {\rm for \ some \ } h  \\
{\rm Im} ( \mu ) < \frac{m-1}{2} & \Longrightarrow  &
\lim_{N \rightarrow \infty} \nu_h =  \mu_1 + i
 \ \ \ \ {\rm for \ some \ } h \ , \\
{\rm Im} ( \mu ) = \frac{m-1}{2} & \Longrightarrow  &
\mu_1, \ldots, \mu_m \mbox{ \ form\ a\ string\ of\ length\ }m \ .
\end{array}
$$
As a consequence,  if $\{ \mu_1, \ldots , \mu_m \}$ is a
subcollection of $\{ \lambda_j \}_{j=1}^l$ which is maximal
among chains of the form (\req(muk)), it must be
a string of length $m$. Hence $\{ \lambda_j \}_{j=1}^l$ is an
union
of reals roots and a finite number of strings of length
greater than one in large $N$ limit. Therefore we obtain
the following conclusion.
\par \vspace{0.2mm} \noindent
{\bf Proposition 2 .} Let $\{ \lambda_j \}_{j=1}^l$ be a
set of Bethe roots for site $N$. As $N \rightarrow \infty$,
$\{ \lambda_j \}_{j=1}^l$ is
composed of real roots toghter with a finite number of strings with length
greater
than one.
$\Box$ \par \vspace{0.2mm} \noindent
{\bf Remark.} The centers of strings in a set of Bethe roots
are indeed all distinct, i.e.
Pauli principle holds for Bethe vectors, for the argument see
e.g. \cite{IK} .

\section{ Bethe Ansatz Equation for a Finite Site $N$ }
In this section, we discuss the Bethe
structure for a finite size $N$.

{\bf Example 1. } Bethe vectors for $l = 1$.
BAE (\req(BAE1)) is described the relation:
\[
 \frac{ \lambda + i/2}{ \lambda - i/2 }  = \omega^k \ , \ \
\ \ \ 1 \leq k \leq N-1 \ , \ \ \ \omega := e^{ 2\pi i/N} \ ,
\]
hence
\[
\lambda = \cot \frac{\pi k}{N} \ , \ \ \ 1 \leq k \leq N-1 \ .
\]
By (\req(ADCBvacuum)) , we have
\[
B ( \lambda ) \Omega_N = i ( \lambda - \frac{i}{2} )^{N-1} \sum_{n=1}^N
\omega^{kn}
(1 \otimes \cdots \otimes \sigma^{-}_{n th}
\otimes \cdots \otimes 1) \Omega_N \ , \ \ \ \mbox{for}
\ \lambda = \cot \frac{\pi k}{N} \ .
\]
Hence the Bethe
state $\Psi_N ( \cot \frac{\pi k}{N} )$ is a multiple
of the vector
\[
\sum_{n=1}^N \omega^{kn} |+> \otimes \cdots |+> \otimes
|->_{n th} \otimes |+> \otimes \cdots |+>
\]
for $ 1 \leq k \leq N-1$, and all these Bethe 1-vectors form
a subspace of $\HZ_N$ of dimension $N-1$.
Note that the above vector for $k = 0$
corresponds to $\lambda = \infty$, which is the solution of
$\frac{ \lambda + i/2}{ \lambda - i/2 } = 1$.
$\Box$ \par \vspace{0.2mm}

{\bf Example 2. } Bethe solutions for $l = 2$. Consider the
change of variables:
\be
z = \frac{ \lambda + \frac{i}{2} }{ \lambda -
\frac{i}{2} } \ \ , \ \ \  \ \
\lambda = \frac{i}{2} \frac{ z + 1 }{ z - 1 } \ .
\ele(zlambda)
We have
\[
\frac{1}{ \overline{z}} = \frac{ \overline{ \lambda } +
\frac{i}{2} }{ \overline{ \lambda } - \frac{i}{2} } \ ,
\]
and
\[
\lambda \in \RZ \cup \{ \infty \} \Longleftrightarrow
| z | = 1 \ .
\]
The $l=2$ BAE becomes
\[
z_1^N = - \frac{ z_1z_2 - 2 z_1  + 1}{z_1z_2 - 2 z_2 + 1 }  \ , \ \
z_1^N z_2^N = 1 \ ,
\]
which is equivalent to
\[
\left\{ \begin{array}{lll}
z_1 z_2 = \omega^k \ \ , & \omega = e^{ 2 \pi i / N} \ , &
\ 0 \leq k \leq N-1 , \\
z_1^{N - 1} = -  \frac{ \omega^k - 2 z_1  + 1}{\omega^k z_1
 - 2 \omega^k + z_1 } \ . &  &
\end{array}  \right.
\]
Note that
if $( z_1 , z_2 )$ is a solution of the above $k$-th equation, so are
$( z_2 , z_1 )$, and $( \frac{1}{ \overline{z_1}} ,
\frac{1}{ \overline{z_2}} )$. Hence both $z_1, z_2,
\frac{1}{ \overline{z_1}}, \frac{1}{ \overline{z_2}} $
are solutions of the equation
\[
( \omega^k + 1 ) z^N - 2 \omega^k z^{N-1} - 2 z
+ ( \omega^k + 1 ) = 0
\]
for some $k$.
Now we are going to determine its solutions. Set
\[
e^{ i \rho } = \omega^{ - k/2 } z \ ,
\]
then
\[
\frac{\omega^k}{z} \
\ \leftrightarrow -\rho \ .
\]
The above equation becomes
\[
(-1)^k \cos \frac{k \pi }{N} e^{ i N \rho }
-  (-1)^k e^{ i (N - 1 ) \rho } -  e^{ i \rho } +
\cos \frac{k \pi }{N} = 0 ,
\]
i.e.
\be
 f_k ( \rho ) : = \frac{ (-1)^k e^{ i (N - 1 ) \rho } +  e^{ i \rho } }
{(-1)^k e^{ i N \rho }+ 1 } = \cos \frac{k \pi }{N} .
\ele(fkcos)
For $k$ odd or $0$, $\rho = 0, \pi$ satisfy
the above equation, and they
correspond to the solutions of BAE with $z_1 = z_2$
, hence not in our consideration.
Note that the function $f_k ( \rho )$ depends only on the
parity of $k$, which can also be expressed by
\[
f_k ( \rho ) =
\left\{ \begin{array}{ll}
\frac{ \sin  ( ( N - 2 ) \rho / 2 ) }
{ \sin(  N \rho / 2 ) } & \mbox{for \ odd \ } k  \\[2mm]
\frac{  \cos  ( ( N - 2 ) \rho / 2 ) }
{ \cos ( N \rho / 2 ) } & \mbox{for \ even \ } k  \ .
\end{array} \right.
\]
Hence $f_k ( \rho )$ has the period $ 2 \pi$ with the symmetries
$f_k ( - \rho ) = f_k ( \rho ), f_k ( \pi - \rho ) = -
f_k ( \rho )$. So it suffices to consider
the solutions with $ 0 \leq \mbox{Re} ( \rho ) \leq \pi$
and $\mbox{Im} ( \rho ) > 0 $ . First let us determine
the real solutions of (\req(fkcos)) for $ 0 < \rho < \pi$.
Claim: $f_k^\prime ( \rho ) \geq 0 $. \par \noindent
Since
\[
f_k^\prime ( \rho ) =
\left\{ \begin{array}{ll}
\frac{ 1 }{ 2 \sin^2(  N \rho / 2 ) }
[ ( N-1) \sin \rho -  \sin  ( N - 1 ) \rho   ]
 & \mbox{for \ odd \ } k  \\[2mm]
\frac{ 1 }{ 2 \cos^2(  N \rho / 2 ) }
[ ( N-1) \sin \rho +  \sin  ( N - 1 ) \rho   ]
 & \mbox{for \ even \ } k  \ ,
\end{array} \right.
\]
it suffices to consider the region of $\rho$ with
\[
\sin  \rho  < \frac{1}{N-1} \ , \ \ \ \ 0 \leq \rho < \pi \ ,
\]
which implies
\[
0 \leq \rho \leq \frac{2}{N-1} \ \ \
\mbox{or} \ \ \ \ \pi - \frac{2}{N-1} \leq \rho \leq \pi \ .
\]
By $f_k ( \pi - \rho ) = -
f_k ( \rho )$, it needs only to consider the region
\[
0 \leq \rho \leq \frac{ \pi }{N} \ ,
\]
hence $f_k^\prime ( \rho ) \geq 0$ for even $k$.
For odd $k$, we have
\[
\frac{d}{d \rho } [ ( N-1) \sin \rho -
\sin  ( N - 1 ) \rho   ]  =  ( N - 1 )
[ \cos \rho - \cos (N-1) \rho ] \geq 0 \
\]
hence
\[
( N-1) \sin \rho -  \sin  ( N - 1 ) \rho  \   \geq \  0 \ ,
\]
which implies $f_k^\prime ( \rho ) \geq 0$. Therefore
$f_k ( \rho )$ is an increasing function on each
connected component of Domain($f_k$), which takes
all values from $- \infty$ to $\infty$. We have
\[
\mbox{Domain} ( f_k )  =
\left\{ \begin{array}{ll}
[ 0 , \pi ] - \{ \frac{ 2j \pi }{N}
\}_{j=1}^{\frac{N-2}{2}} & \mbox{for \ odd \ } k  \\[2mm]
[ 0 , \pi ] - \{ \frac{ (2j-1) \pi }{N}
\}_{j=1}^{\frac{N}{2}} & \mbox{for \ even \ } k  \ .
\end{array} \right.
\]

\begin{figure}
\vspace*{-10mm}
\begin{center}
\caption{\label{fig1}
         $ k$ : odd , $f_k ( \rho ) =
\frac{ \sin  ( ( N - 2 ) \rho / 2 ) }
{ \sin(  N \rho / 2 ) }
\ , \ 0 \leq \rho \leq \pi , \ $  with $N = 20$}
\end{center}
\end{figure}

\begin{figure}
\vspace*{-10mm}
\begin{center}
\leavevmode\epsfxsize=120mm \epsffile{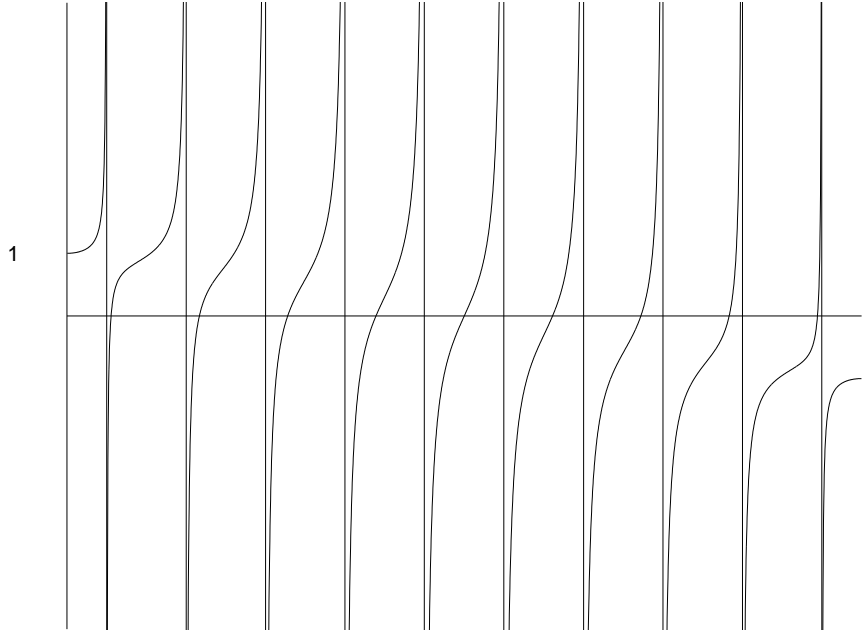}
\caption{\label{fig2}
        $ k$ : even , \ $f_k ( \rho ) =
 \frac{  \cos  ( ( N - 2 ) \rho / 2 ) }
{ \cos ( N \rho / 2 ) } \ , \ 0 \leq \rho \leq \pi , \ $  with $N= 20$  }
\end{center}
\end{figure}
Hence the number of real solutions $\rho$  of (\req(fkcos))
with $ 0 < \rho < \pi$  is given by
\[
\left\{ \begin{array}{ll}
\frac{N}{2}- 2 &
\mbox{for \ odd \ } k \ \mbox{and \ } \  | \cos \frac{k \pi}{N} |
< \frac{N-2}{N} \ , \\
\frac{N}{2}- 1 &
\mbox{for \ odd \ } k \ \mbox{and \ } | \cos \frac{k \pi}{N} |
\geq \frac{N-2}{N} \ ,  \\
\frac{N}{2} - 1  & \mbox{for \ even \ } k  \ .
\end{array} \right.
\]
The total number is
\[
\frac{N}{2} ( N - 3 ) + \# \{ k = 2j - 1 \ | \ 1 \leq j \leq
\frac{N}{2} \ , \
| \cos \frac{k \pi}{N} | \geq \frac{N-2}{N} \} \ ,
\]
and it is equal to the number of real solutions $\{ \lambda_1 ,
\lambda_2 \} $ of BAE for $l = 2$, including those
with the value $\infty$, which is the solutions corresponding to
$\rho = \frac{ k \pi }{N}$
for the equation (\req(fkcos)).
Hence the number of finite
real Bethe solutions $\{ \lambda_1 , \lambda_2 \} $ for
$l = 2$ is equal to
\be
\frac{N}{2} ( N - 3 ) - ( N - 1 ) + \# \{ k = 2j - 1 \ | \ 1 \leq j \leq
\frac{N}{2} \ , \
| \cos \frac{k \pi}{N} | \geq \frac{N-2}{N} \} \ .
\ele(l=2real)
Since the total number of complex solutions of (\req(fkcos)) is
$N$, one obtains the contribution
of non-real Bethe solutions for $l = 2$ whose number
is given by
\[
\left\{ \begin{array}{ll}
 1 &
\mbox{for \ odd \ } k \ \ \ \mbox{and \ } | \cos \frac{k \pi}{N} |
< \frac{N-2}{N} \ , \\
 0 &
\mbox{for \ odd \ } k \ \ \ \mbox{and \ } | \cos \frac{k \pi}{N} |
\geq \frac{N-2}{N} \ ,  \\
1  & \mbox{for \ even \ } k \ \ \ \mbox{and \ } k \neq 0 \ , \\
0  & \mbox{for \  } k = 0 \ ,
\end{array} \right.
\]
with the total number
\be
N - 1 - \# \{ k = 2j - 1 \ | \ 1 \leq j \leq
\frac{N}{2} \ , \
| \cos \frac{k \pi}{N} | \geq \frac{N-2}{N} \} \ .
\ele(l=2comp)
Therefore the number of Bethe solutions for $l=2$ is
 the sum of (\req(l=2real)) and (\req(l=2comp)), which is equal
to $C^N_2 - C^N_1$. This coincides with the counting of
Bethe 2-states given in \cite{TF81} .
$\Box$ \par \vspace{0.2mm}

{\bf Example 3. } Bethe solutions for $ l = 3 \ , N = 6 $. The
BAE is given by
\[
( \frac{ \lambda_j + \frac{i}{2} }{ \lambda_j - \frac{i}{2} }
)^6 =
- \prod_{m=1}^{3} \frac{ \lambda_j - \lambda_m + i
}{ \lambda_j - \lambda_m - i} \ \
\ \ \ ( j = 1 , 2, 3 ) \
\ , \ \lambda_j \neq \lambda_k \in \CZ  \ .
\]
Claim: There are exactly 5 solutions of the above equation,
and each of them is invaraint under the map
$\lambda \rightarrow - \lambda$.

First let us consider a solution invariant under the sign
symmetry:
$\lambda \leftrightarrow - \lambda$. We may assume $\lambda_j$'s
take the form
\[
\lambda_1 = - \lambda_3 = \lambda \ \ ( \neq 0 ) \ , \ \ \
\lambda_2 = 0 \ .
\]
With the variable $z$ in (\req(zlambda)), the corresponding
equation of $z_1$ is
\[
z^6 - 3 z^5 + 3 z - 1 = 0 \ .
\]
The above equation has 6 distinct roots, which includes
$z = 1$.
The value $\lambda$ corresponding to each of the other
5 solutions gives a Bethe solution symmetric under the change
of sign. They contribute 5 independent states in the Hilbert
space $\HZ_6$, which is a  64-dimensional vector space .
On the other hand, by Example 1 and 2 of this section together
with (iv) of Proposition 1, the total number of Bethe states
$\Psi_N ( \lambda_1, \ldots , \lambda_l ) $ for $0 \leq l \leq
2$
is equal to
\[
1 \times 7 + 5 \times 5 + 9 \times 3 = 59 \ .
\]
Therefore there is no other Bethe 3-state except the symmetric
ones we have described. Then the conclusion follows
immeadiately.
$\Box$ \par \vspace{0.2mm}

Explicit Bethe solutions of a higher $l$
are difficult to obtain in general for a finite size
$N$. For the rest of
this paper, we shall consider only the case
\[
l  = g : = \frac{N}{2} \ \ {\rm and } \ \ \ \lambda_j \in \RZ
\ \ \ {\rm for } \ \  j= 1, \ldots , g \ .
\]
The equation we are going to discuss is the following
form:
\be
( \frac{ \lambda_j + \frac{i}{2} }{ \lambda_j - \frac{i}{2} } )^N =
- \prod_{m=1}^{g} \frac{ \lambda_j - \lambda_m + i
}{ \lambda_j - \lambda_m - i} \ \ , \ \ \lambda_j \in \RZ  \ \
( j = 1 , \ldots, g ) \ .
\ele(BAEgs)
The Bethe vector corresponding to the above Bethe roots
leads to the ground state of antiferromagetic $H_{\rm XXX}$
 (i.e. $J < 0 $ ) in $N \rightarrow \infty$.
It is more convenient to consider the logarithmic form of the
the above equation. By using the relation
\[
\frac{1}{i} \log \frac{ \lambda - i}{\lambda + i} =
2 \arctan( \lambda ) + \pi \ , \ \ \ \
| \arctan( \lambda ) | < \frac{\pi}{2}  \ \ \ \
 \mbox{ for \ } \lambda \in \RZ ,
\]
$$
\put (-50,25 ){ \line(1, 0){80}}
\put (-45,0 ){ \line(0, 1){50}}
\put (-45,5){\vector( 3 ,1){58}}
\put (-45,45){\vector( 3, -1){58}}
\put (-40,0){\shortstack{$-i$}}
\put (-40,45){\shortstack{$i$}}
\put(13, 25){\vector( 3 ,1){30}}
\put(13, 25){\vector( 3 ,-1){30}}
\put(13, 30){\shortstack{$\lambda$}}
\put( -90, 35){\shortstack{$\arctan \lambda$}}
\put(-50,38){\vector(3,1){12}}
$$
we obtain the following equation from (\req(BAEgs)):
\[
\arctan 2 \lambda_j  = \pi \frac{Q_j}{N} + \frac{1}{N}
\sum_{k=1}^{ g }
\arctan ( \lambda_j - \lambda_k ) \ \ \ , \ 1 \leq j
\leq g \ ,
\]
where $Q_j$'s are all integers or all half-integers according
to $g$ odd or even, and they are bounded by the
number $\widetilde{Q}$ determined by the relation:
\[
\arctan ( \infty ) = \frac{\pi}{N}(\widetilde{Q} + \frac{1}{2}) +
\frac{1}{N} \sum_{k=1}^{ g }
\arctan ( \infty ) \ .
\]
Hence
$$
\begin{array}{rl}
\widetilde{Q} & =  \frac{N}{4} - \frac{1}{2} \\
q_j :  = \frac{Q_j}{N} & = \frac{-1}{4} - \frac{1}{2N} + \frac{j}{N} \ , \ \
\ \ j = 1, 2, \dots, g  ,
\end{array}
$$
and (\req(BAEgs)) is now equivalent to
the equation
\bea(c)
F_j ( \Lambda ) : =
\arctan 2 \lambda_j  - ( \pi q_j + \frac{1}{N}
\sum_{k=1}^g
\arctan ( \lambda_j - \lambda_k ) ) = 0 \\ [2mm]
\Lambda : = \left( \begin{array}{c}
 \lambda_1 \\
 \vdots  \\
\lambda_g
\end{array} \right) \in \RZ^g\ ,
\ \ j = 1, 2, \dots, g.
\elea(gslog)
Since only a finite number of solutions can be obtained for BAE
(\req(BAEgs)) by Propostion 1, we obtain the following result.
\par \vspace{0.2mm} \noindent
{\bf Lemma 2 .} BAE (\req(BAEgs)) is equivalent to the
equation (\req(gslog)) , which has at most a finite number of
solutions.
$\Box$ \par \vspace{0.2mm} \noindent

Now we are going to show the existence of real solutions for
(\req(gslog)).
Define the endomorphism $F$ and the linear involution
$E$ of $\RZ^g$ by
$$
\begin{array}{l}
F : \RZ^g \longrightarrow \RZ^g \ , \ \
F ( \Lambda  ) :=
\left( \begin{array}{c}
 F_1 ( \Lambda   ) \\
 \vdots  \\
F_g ( \Lambda   )
\end{array} \right) \\
E : \RZ^g \longrightarrow \RZ^g \ , \ \
E ( \Lambda ) =
\left( \begin{array}{c}
 \vdots  \\
E ( \Lambda )_j \\
\vdots
\end{array} \right) \ , \ \
E ( \Lambda )_j := -\lambda_{g+1-j} \ .
\end{array}
$$
Denote the $E$-invariant vector in $\RZ^g$:
\[
q = \left( \begin{array}{c}
 q_1  \\
 \vdots  \\
q_g
\end{array} \right) .
\]
It is easy to see that $F$ has the following symmetry
properties:
\par \vspace{0.2mm} \noindent
{\bf Lemma 3. }

(i)
\[
 F ( - \Lambda ) = - F ( \Lambda ) - 2 \pi q \ \ \ \
{\rm for \ } \ \Lambda \in \RZ^g \ .
\]

(ii) $F \circ E = E \circ F $.
$\Box$ \par \vspace{0.2mm} \noindent
We are going to show the existence of solutions of the equation
(\req(gslog)) by the fixed point theory.
\par \vspace{0.2mm} \noindent
{\bf Proposition 3 . } For a sufficient large cube $C$ in $\RZ^g$
\[
C = \{ \Lambda = ( \lambda_1, \ldots, \lambda_g )^{t} \in
\RZ^g : -a \leq x_j \leq a , \ \ 1 \leq j \leq g \ \}
\]
there exists a solution for the equation
 $F ( \Lambda ) = \vec{0}$ in $\Lambda \in C$.
\par \vspace{0.2mm} \noindent
{\it Proof. } Since
\[
\lim_{ x \rightarrow \infty } ( \arctan 2x - \frac{\pi}{4} )
= \frac{\pi}{4}
\]
there is a positive number $\alpha$ such that
\[
\arctan 2\alpha - \pi q_g = \arctan 2\alpha - \frac{\pi}{4} + \frac{\pi}{2N}
> \frac{\pi}{4} \ .
\]
Let $C$ be the cube in $\RZ^g$ with
\[
a \geq \alpha \ \ .
\]
Denote $C_j^+ , C_j^-$ the faces of $C$ :
\[
C_j^+ = \{ \Lambda \in C \ | \ \lambda_j = a \} \ \ , \ \ \
C_j^- = \{ \Lambda \in C \ | \ \lambda_j = -a \} \ \ , \ \ \
j = 1, \ldots , g \ .
\]
By the inequalities
\[
\arctan 2 a - \pi q_j  \geq \arctan 2 \alpha - \pi q_g >
\frac{\pi}{4} \ \ ,
\ \ \ \
\frac{1}{N} \sum_{k=1}^g
\arctan ( a - \lambda_k ) )  \leq \frac{g \pi }{2 N}
= \frac{\pi}{4}
\]
one has
\[
F_j ( \Lambda ) > 0 \ \ {\rm for } \ \ \ \ \Lambda \in
C_j^+ \ .
\]
For $\Lambda \in C_j^-$ and $j > [ \frac{N}{4} ]$,
we have
\[
q_j \geq 0 \ , \ \ - \Lambda \in C_j^+
\]
hence by (i) of Lemma 3,
\[
F_j ( \Lambda ) = - F_j ( - \Lambda ) - 2 \pi q_j < 0 \ .
\]
For $\Lambda \in C_j^-$ and $j \leq [ \frac{N}{4} ]$,
we have
\[
E ( \Lambda ) \in C_{g+1-j}^+ \ , \ \
F_{g+1-j} ( E ( \Lambda ) ) > 0
\]
hence by (ii) of Lemma 3,
\[
F_j ( \Lambda ) = - F_{g+1-j} ( E ( \Lambda ) ) < 0 \ .
\]
Therefore we obtain
\[
F_j ( \Lambda ) < 0 \ \ \ \ \ {\rm for } \ \Lambda \in
C_j^- \ .
\]
Thus by Poincar\'{e}-Miranda fixed point theorem,
(see e.g. \cite{Brow} p.p.12), the conclusion of the proposition follows
immediately.
$\Box$ \par \vspace{0.2in} \noindent
{\bf Remark.} (i) One can require the symmetry property on
the solutions in the above proposition. Indeed there is a
solution of $F( \Lambda ) = 0$
in the intersect of cube $C$ with the hypersurface $H$,
\[
H : = \{ \Lambda \in \CZ^g \ | \ E ( \Lambda ) =
\Lambda \} \ .
\]
In fact, the map $F$ sends $H$ into itself. Since
$\lambda_j , 1 \leq j \leq [ \frac{N}{4}] ,$ form a coordinate
system of $H$, the conclusion in the above proof implies that the map
\[
F_{\rm rest} : H \cap C \longrightarrow H
\]
also satisfies the the conditions of
Poincar\'{e}-Miranda fixed point theorem, hence it follows the
result.

(ii) The argument given in the proposition also
provides an appriori estimate on the location of roots of
$F ( \Lambda ) = \vec{0}$, which lies in the following region
\[
\{ \Lambda \ | \ b_j \leq \lambda_j \leq a_j \ ,
\ 1 \leq j \leq g \ \}
\]
where
\[
a_j = \frac{1}{2} \tan ( \frac{\pi}{4} + \pi q_j) \ \ \ \ \
b_j = \frac{1}{2} \tan ( \frac{-\pi}{4} + \pi q_j) \ \  .
\]
$\Box$ \par \vspace{0.2mm} \noindent
As a corollary of Proposition 3 and Lemma 2 , we have
the following result:
\par \vspace{0.2mm} \noindent
{\bf Proposition 4 . } There is a solution of the equation
(\req(BAEgs)).
$\Box$ \par \vspace{0.2mm} \noindent
The uniqueness for the solution of (\req(BAEgs)) should
be expected by the thermodynamic nature of the solutions.
Let us look a few cases of small $N$.
Note that if $\{ \lambda_j \}_{j=1}^g$ satisfies the equation
(\req(gslog)), they satisfy the following equality:
\[
\sum_{j=1}^g \arctan (2 \lambda_j ) = 0 \ .
\]
For $N=4$, the above symmetry relation enables one to obtain the
solution of the equation (\req(gslog)) which is equivalent to
\[
\left\{ \begin{array}{l}
 \lambda _1 + \lambda_2 = 0 \  \\
\arctan ( 2 \lambda _1)  =  \frac{- \pi}{8} +
\frac{1}{4} \arctan ( \lambda_1 - \lambda_2 )
\end{array} \right.
\]
Hence
\[
\lambda_1 =  \frac{-1}{2 \sqrt{3} } \ , \ \ \
\lambda_2 =  \frac{1}{2 \sqrt{3} } \ .
\]
For a general $N$, one can conclude that all the $\lambda_j$'s can
not be of the same sign.
Indeed one can determine signs of $\lambda_1$ and $\lambda_g$
from the first and the last relations in (\req(gslog)):
\[
\arctan ( 2 \lambda_1 ) < \pi (  \frac{-1}{4} + \frac{1}{2N} ) +
( \frac{N}{2} - 1 ) \frac{\pi}{2N} = 0 \ , \ \
\arctan ( 2 \lambda_g ) > \pi (  \frac{1}{4} - \frac{1}{2N} )
- ( \frac{N}{2} - 1 ) \frac{ \pi}{2N} = 0
\]
hence
\be
\lambda_1 < 0 < \lambda_g \ .
\ele(1g)
It appears to be the case that certain
symmetry properties exist among $\lambda_j$'s, e.g.
\[
\lambda_1 < \lambda_2 < \cdots < \lambda_g \ \ , \ \ \ \ \ \
\lambda_j + \lambda_{g+1-j} = 0 \ ,
\]
but the mathematical derivation from the equation
(\req(gslog)) seems a difficult problem, even in the case
of $N = 6$:
\be
\left\{ \begin{array}{lll}
\arctan ( 2 \lambda _1) & = & \frac{-\pi}{6} +
\frac{1}{6} ( \arctan ( \lambda_1 - \lambda_2 ) +
\arctan ( \lambda_1 - \lambda_3 ) ) \\
\arctan ( 2 \lambda _2) & = &
\frac{1}{6} ( \arctan ( \lambda_2 - \lambda_1 ) +
\arctan ( \lambda_2 - \lambda_3 ) ) \\
\arctan ( 2 \lambda _3) & = & \frac{\pi}{6} +
\frac{1}{6} ( \arctan ( \lambda_3 - \lambda_1 ) +
\arctan ( \lambda_3 - \lambda_2 ) )
\end{array} \right.
\ele(N6)
However in the above case, by the analysis of Example
3 in this section, the symmetry property for the solutions
holds:
\[
\lambda_1 + \lambda_3 = 0 \ \ , \ \ \ \lambda_2 = 0 \ .
\]
Hence $\lambda_1$ satisfies the equation
\[
5 \arctan ( 2 \lambda ) - \arctan ( \lambda ) = \pi \ .
\]
Since the left hand side is a strictly inceasing function of
$\lambda$, there exists an unique solution of
$\lambda_1$, hence the uniqueness of the equation (\req(N6)).
For a larger size $N$, the mathematical structure of the
equation (\req(gslog)) becomes more complicated that no effective
mean could be found at this moment for the uniqueness problem.
In the next section, we shall present an plausible,
but not a mathematically  rigorous, argument on
the unique ground state solution for a large but
finite $N$ based on the thermodynamic limit procedure.

\section{Ground State for Antiferromagnetic XXX Model }
The ground state of the Hamiltonian $H_{
\mbox{XXX} }$ for the
antiferromagnetic case is the state for the solution of
(\req(gslog)).
In the thermodynamic limit, one assumes there is a
real solution $\{ \lambda_j \}_{j=1}^{\frac{N}{2}}$
for the asymtotic equation of (\req(gslog)):
\[
\arctan 2 \lambda_j  \sim \pi q_j + \frac{1}{N}
\sum_{k=1}^{ \frac{N}{2} }
\arctan ( \lambda_j - \lambda_k ) \ , \ \ 1 \leq j
\leq \frac{N}{2} \ .
\]
The $q_j$'s are considered as quantum numbers of the ground
state.
As $N$ tends to $\infty$, the continuous
version of the above relation is obtained by the following
substitution :
\[
q_j \longrightarrow x \in
( \frac{-1}{4}, \frac{1}{4} ) \ \ , \ \ \ \
\lambda_j \longrightarrow \lambda ( x ) \ \ ,
\]
here $\lambda( x )$ is a monotonic increasing function with
$\lambda(\frac{-1}{4} ) = - \infty $ and $
\lambda(\frac{1}{4} ) = \infty $. The density of the ground state is now
defined to be
\[
\rho_v ( \lambda ) = \frac{ d x } { d \lambda } \ \ \
( \ = \lim_{ N \rightarrow \infty,
\lambda_j \rightarrow \lambda }
\frac{1}{N ( \lambda_{j+1} - \lambda_j )} \ ) .
\]
The logarithmic BAE  for the ground state
now becomes
\[
\arctan 2 \lambda ( x ) = \pi x + \int_{ -\frac{1}{4} }^{\frac{1}{4}}
\arctan ( \lambda ( x ) - \lambda ( y ) ) dy .
\]
Differentiating the above equation with respect to $x$,
one can derive the integral equation for $\rho_v$:
\be
\pi \rho_v ( \lambda)+
( \frac{1}{ 1 + \mu^2 } \ast \rho_v ) ( \lambda )
= \frac{2}{1+4 \lambda^2} \
\ele(denvXXX)
here the convolution of function $f$ and $g$ is defined by
\[
( f \ast g ) ( \lambda ) = \int_{ -\infty }^{\infty}
f ( \lambda - \mu ) g ( \mu ) d \mu .
\]
Then the density, energy , momentum and spin of the ground
state in the thermodynamic limit for the antiferromagnetic
XXX model are given by
\[
\rho_v ( \lambda ) = \frac{1}{ 2 \cosh ( \pi \lambda ) } , \ \
E_v =  J N \log 2 \ , \ \
\Pi_v = \frac{ N \pi}{2} \ \ ( \mbox{mod \ } 2 \pi ) , \ \
S_v = 0 \ .
\]
( For the details, see \cite{FT} \cite{YY} ).

Now we explain the reason for the uniqueness of the ground
state for a large but finite $N$.
Suppose that $\{ \lambda_j \}_{j=1}^{N/2}$ is a set of
solutions of (\req(gslog)) such that the density $\rho_v ( \lambda )$
of the continuous limit $\lambda ( x )$ is given by
$\frac{1}{ 2 \cosh ( \pi \lambda ) } $. Without loss of
generality, we may assume there is another set of real
solutions $\{ \widetilde{\lambda}_j \}_{j=1}^l$ of the equation
(\req(gslog)) for a size $N$. Then for some $\mu_j$, one has the
expansion:
\[
\widetilde{\lambda}_j = \lambda_j + \frac{1}{N} \mu_j +
O ( \frac{1}{N^2} ) \ \ \ \ \ , \
j = 1, \ldots , \frac{N}{2}  \ .
\]
We have
\[
\arctan ( 2 \lambda_j ) + \frac{2 \mu_j}{N ( 1+4\lambda_j^2 )}
 = \pi q_j + \frac{1}{N} \sum_{k=1}^{N/2}
[ \arctan ( \lambda_j - \lambda_k ) +
\frac{1}{1+ ( \lambda_j - \lambda_k )^2} \frac{ \mu_j -\mu_k}{N}
] + O ( \frac{1}{N^2} ) \ ,
\]
hence
\[
\frac{2 \mu_j}{N ( 1+4\lambda_j^2 )}  =
 \frac{1}{N} \sum_{k=1}^{N/2}
[ \arctan ( \lambda_j - \lambda_k ) +
\frac{1}{1+ ( \lambda_j - \lambda_k )^2} \frac{ \mu_j -\mu_k}{N}
] + O ( \frac{1}{N^2} ) \ .
\]
In the continuous limit,
\[
\lambda_j \longrightarrow \lambda ( x ) \ , \ \
\mu_j \longrightarrow \mu ( x ) \ \ \ \ {\rm for } \ \ x \in (
\frac{-1}{4}, \frac{1}{4} ) \ ,
\]
the above equation becomes
\[
\frac{2 \mu ( x )}{ 1+4 \lambda ( x )^2 } =
\int_{\frac{-1}{4}}^{\frac{1}{4}}
\frac{\mu (x) -\mu (y)}{1+ ( \lambda (x) - \lambda (y) )^2} dy
\ .
\]
Therefore
\[
\frac{2 \mu ( \lambda )}{ 1+4 \lambda^2 }  =
\int_{-\infty}^{\infty}
\frac{\mu (\lambda ) -\mu (\nu)}{1+ ( \lambda - \nu )^2}
 \rho_v (\nu) d\nu \
\]
and
\[
\mu ( \lambda ) [
\frac{2 \mu ( \lambda )}{ 1+4 \lambda^2 } -
\int_{- \infty}^{\infty}
\frac{\rho_v (\nu)}{1+ ( \lambda - \nu )^2} d\nu ]
 = - \int_{-\infty}^{\infty}
\frac{ \mu ( \nu ) \rho_v (\nu)}{1+ ( \lambda - \nu )^2} d\nu .
\]
By (\req(denvXXX)), one has
\[
\pi \mu ( \lambda ) \rho_v (\lambda) =
- \int_{-\infty}^{\infty}
\frac{ \mu ( \nu ) \rho_v (\nu)}{1+ ( \lambda - \nu )^2} d\nu \
\ .
\]
By taking Fourier transform, the above equation becomes
\[
\pi \widehat{ \mu \rho_v } =
\widehat{\frac{-1}{1+\lambda^2}} \widehat{ \mu \rho_v }
\]
which implies $\mu ( x ) = 0$. Therefore $\lambda_j$ and
$\widetilde{\lambda}_j$ agree up to the order of
$\frac{1}{N^2}$. Repeating the same procedure inductively
to higher order terms of $\frac{1}{N}$, one arrives the
conclusion that
$\lambda_j$ coincides with $\widetilde{\lambda}_j$ for all $j$.


\begin{thebibliography}{99}
\bibitem{B} R. J. Baxter, Exactly solved models in statistical mechanics,
Academic Press (1982).
\bibitem{Brow} F. E. Browder, Fixed point theory and
non-linear problems, Bull. AMS 9 (1983) 1-39.
\bibitem{EKK} F. H. L. Essler, V. E. Korepin and K. Schoutens,
Fine structur of the Bethe ansatz for the spin-$\frac{1}{2}$ Heisenberg
XXX model, J. Phys. A: Math. Gen. 25 (1992) 4115-4126.
\bibitem{F} L. D. Faddeev, Algebraic aspects of Bethe-Ansatz, ITP-SB-94-11,
\bibitem{FT} L. D. Faddeev and L. A. Takhtadzhan , Spectrum
and scattering of excitations in the one-dimensional
isotropic Heisenberg model, J. Soviet Math. 24 (1984) 241-267.
\bibitem{IK} A. G. Izegin and V. E. Korepin, Pauli principal
for one-dimensional bosons and the algebraic Bethe Ansatz,
Lett. Math. Phy. 6 (1982) 283-288.
\bibitem{Ta} M. Takahashi , Prog. Theor. Phys. 48 (1971) 401 .
\bibitem{TF79} L. A. Takhtadzhan and L. D. Faddeev , The quantum method
of the inverse problem and the Heisenberg XYZ model, Russian Math. Surveys
34: 5 ( 1979 ), 11 - 68.
\bibitem{TF81} L. A. Takhtadzhan and L. D. Faddeev , What is
the spin of a spin wave, Phys. Letts 85A, No 6,7 (1981)
375 - 377.
\bibitem{YY} C. N. Yang and C. P. Yang, One dimensional chain
of anisotropic spin-spin interaction. I. Proof of Bethe's
hypothesisfor ground state in finite system, Phy. Rev. 150
(1966) 321-327 , II. Properties of ground state energy per
lattice site for a finite system, Phy. Rev. 150 (1966)
327-339.
\end{thebibliography}
\end{document}